\documentclass[12pt]{article}
\usepackage{graphicx}
\usepackage[cp1251]{inputenc}
\usepackage{epsfig}
\usepackage[english]{babel}

\date{}

\begin{document}
\setcounter{page}{1}
\pagestyle{plain}

\title{\bf{Controllable intrinsic DC spin/valley Hall conductivity in ferromagnetic silicene:
Exploring a fully spin/valley polarized transport}}

\author{Yawar Mohammadi$^1$\thanks{Corresponding author. Tel./fax: +98 831 427
4569, Tel: +98 831 427 4569. E-mail address:
yawar.mohammadi@gmail.com}  and Borhan Arghavani Nia$^2$}
\maketitle{\centerline{$^1$Young Researchers and Elite Club,
Kermanshah Branch, Islamic Azad University, Kermanshah,
Iran}\maketitle{\centerline{$^2$Department of Physics, Kermanshah
Branch, Islamic Azad University, Kermanshah, Iran}

\begin{abstract}

We study intrinsic DC spin and valley Hall conductivity in doped
ferromagnetic silicene in the presence of an electric filed
applied perpendicular to silicene sheet. By calculating its energy
spectrum and wavefunction and by making use of Kubo formalism, we
obtain a general relation for the transverse Hall conductivity
which can be used to obtain spin- and valley-Hall conductivity.
Our results, in the zero limit of the exchange field, reduces to
the previous results. Furthermore we discuss electrically tunable
spin and valley polarized transport in ferromagnetic silicene and
obtain the necessary conditions for observing a fully spin or
valley polarized transport.

\end{abstract}


\vspace{0.5cm}

\section{Introduction}
\label{sec:01}

Since successful isolation of a single layer of
graphite\cite{Novoselov1}, graphene, as the first real
two-dimensional lattice structure which shows novel appealing
properties\cite{Castro Neto1,Peres1}, many researchers try to
synthesis or isolate new two-dimensional materials. These efforts
result in finding other two dimensional materials such as
BN\cite{Pacile1}, transition metal dichalcogenides
(TMDs)\cite{Novoselov2} and recently a monolayer of silicon, known
as silicene\cite{Lalmi1,Padova1,Aufray1,Vogt1}.

Silicene is a monolayer of silicon atoms arranged in a honeycomb
lattice structure as similar as graphene. While, as in graphene,
its low-energy dynamics near the two valleys at the corner of the
hexagonal Brillouin zone is described by the Dirac theory, its
Dirac electrons, due to a large spin-orbit (SO) interaction, are
massive with a energy gap as $1.55meV$\cite{Liu1,Drummond1}.
Furthermore, due to the large ionic radius, silicene is
buckled\cite{Liu1} such that the A and B sublattices of honeycomb
lattice shifted vertically with respect to each other and sit in
two parallel planes with a separation of $0.46
nm$\cite{Drummond1,Ni1}. The buckled structure of silicene allows
to tune its band gap via an electric filed applied perpendicular
to its layer. These features donate many attractive properties to
silicene
\cite{Drummond1,Ezawa1,Ezawa2,An1,Tahir1,Ezawa3,Tsai1,Tabert1,Pan1}.

The SO interaction in silicene is strong, so it is a suitable
candidate to study the spin-based effects. Due to this fact,
recently silicene has been the subject of strong
interest\cite{Dyrda1,Tahir2,Tabert2,Duppen1,An2}. In addition to
the spin degree of freedom which is the footstone of the
spintronics, the valley degree of freedom in silicene, as in
graphene\cite{Rycerz1,Xiao1,Akhmerov1} and
$MoS_{2}$\cite{Xiao2,Zeng1,Mak1,Cao1}, can be manipulated and
hired in a new technology known as valleytronics. One can populate
states preferentially in one valley to achieve valley
polarization. One way is to use circular polarized light which was
discussed theoretically\cite{Xiao2}. Another way is to apply a
vertical external magnetic filed to silicene sheet, so Landau
levels form in the electronic density of states. Then, if an
excitonic gap via an external vertical voltage included, $n=0$
Landau level splits into distinct valley- and spin-polarized
levels\cite{Tabert1}. This is in contrast to that occurs in
graphene, in which $n=0$ Landau level only splits between two
distinct valley-polarized spin degenerated energy
levels\cite{Gusynin1,Gusynin2,Sadowski1,Jiang1}. In other way, as
in graphene\cite{Haugen1,Tombros1}, the spin/valley polarized
current is obtained in silicene\cite{Yokoyama1} junctions by
deposing a ferromagnet on the top of its surface. These features
make silicene a promising candidate for spin- and valleytronic
technology.

In this paper, we consider DC valley/spin Hall conductivity in a
ferromagnetic silicene (a silicene sheet with ferromagnet deposed
on the top of its surface). We obtain a general relation for its
transverse Hall conductivity which can be use to calculate
spin/valley Hall conductivity and to discuss possible phase
transitions. Furthermore, we obtain the conditions necessary to
realize fully valley/spin polarized transport, which depends on
the doping, exchange magnetization and the applied perpendicular
electric field. The paper is organized as follows. Sec.II is
devoted to introduce the Hamiltonian model and obtain the general
relation for the transverse Hall conductivity. In Sec.III we
present our results and discussion. Finally in Sec.IV we end this
paper by summary and conclusions.

\section{Model Hamiltonian}
\label{sec:02}

The low-energy dynamic in a ferromagnetic silicene, subjected to a
uniform electric field applied perpendicular to sislicene's plane,
is given by\cite{Liu1,Yokoyama1}
\begin{eqnarray}
H_{\eta,s_{z}}=\hbar v_{F}(k_{x}\tau_{x}-\eta k_{y}\tau_{y})-\eta
s_{z}\Delta_{so}\tau_{z}+\Delta_{z}\tau_{z}-s_{z}M, \label{e1}
\end{eqnarray}
which acts in the sublattice pseudospin space with a wavefunction
as
$\Psi^{\eta,s_{z}}=\{\psi_{A}^{\eta,s_{z}},\psi_{B}^{\eta,s_{z}}\}^{T}$.
The first part of the Hamiltonian is the Dirac hamiltonian
describing the low-energy excitations around Dirac points
($\mathbf{K}$ and $\mathbf{K^{'}}$ denoted by $\eta$ index) at the
corners of the hexagonal first Brilouin zone. This term arises
from nearest neighbor energy transfer. $v_{F}=$ is the Fermi
velocity of silicene, $\tau_{i}$ ($i=x,y,z$) are the Pauli
matrixes and $\mathbf{k}=(k_{x},k_{y})$ is the two dimensional
momentum measured from Dirac points. The second term is the
Kane-Mele term for the intrinsic spin-orbit coupling, where
$\Delta_{so}=3.9 meV$\cite{Liu1} denotes to the spin-orbit
coupling and $s_{z}$ index referred to two spin degrees of
freedom, up ($s_{z}=+$) and down ($s_{z}=-$). The third term is
the on-site potential difference between $A$ and $B$ sublattice,
arising from the buckled structure of silicene when a
perpendicular electric field is applied with $\Delta_{z}=E_{z}d$
where $E_{z}$ is the electric field and the $2d=0.46nm$ is the
vertical separation of two different sublattice's plane. The last
term is the exchange magnetization where $M$ is the exchange
field. The exchange field my be due to proximity effect arising
from a magnetic adatom deposed on the surface of the
silicene\cite{Qiao1} or from a magnetic insulator substrate like
$EuO$ as proposed for graphene\cite{Haugen1}.

We obtain the energy spectrum, by diagonalizing the Hamiltonian
matrix given in Eq.(\ref{e1}), as
\begin{eqnarray}
\varepsilon^{\eta,s_{z}}_{\nu}=\nu\sqrt{\Delta_{\eta,s_{z}}^{2}+(\hbar
v_{F}k)^{2}}-s_{z}M, \label{e2}
\end{eqnarray}
where $\nu=+(\nu=-)$ denotes the conduction (valance) bands,
$\Delta_{\eta,s_{z}}=\eta s_{z}\Delta_{so}-\Delta_{z}$ and
$k=\sqrt{k_{x}^{2}+k_{y}^{2}}$. The corresponding wavefunctions
are given by
\begin{eqnarray} \Psi^{\eta,s_{z}}_{\nu}(\mathbf{k})=
\frac{e^{i\mathbf{k}.\mathbf{r}}}{\sqrt{2\chi_{\eta,s_{z}}}}\left(
\begin{array}{c}
\sqrt{\chi_{\eta,s_{z}}-\nu \Delta_{\eta,s_{z}}}  \\
\nu \sqrt{\chi_{\eta,s_{z}}+\nu \Delta_{\eta,s_{z}}} e^{-i\eta
\phi_{k}}
\end{array}
\right),\label{e3}
\end{eqnarray}
where $\chi_{\eta,s_{z}}=\sqrt{\Delta_{\eta,s_{z}}^{2}+(\hbar
v_{F}k)^{2}}$ and $\phi_{k}=\tan^{-1}(k_{y}/k_{x})$. Figure
\ref{f1} shows the energy spectrum of silicene (Fig. \ref{f1}(a))
and ferromagnetic silicene with $M=\Delta_{so}/2$ for three
different values of the electric field, $\Delta_{z}=0$ plotted in
Fig. \ref{f1}(b), $\Delta_{z}=\Delta_{so}$ in Fig. \ref{f1}(c) and
$\Delta_{z}=2\Delta_{so}$ in Fig. \ref{f1}(d). These figures shows
the energy spectrum around $\mathbf{K}$. The energy spectrum
around $\mathbf{K}^{'}$ for zero electrical potential,
$\Delta_{z}=0$, is equal to that of $\mathbf{K}$ point. To obtain
the other energy-spectrum plots it is enough to reflect the
energy-band plots with respect to $E=0$ and exchange spin up and
down.

DC transverse Hall conductivity, $\sigma_{xy}$, written in the
Kubo formalism, is given by\cite{Gusynin3,Vargiamidis1}
\begin{eqnarray}
\sigma^{\eta,s_{z}}_{xy}=&&-i\frac{e^{2}\hbar}{A}\sum_{\mathbf{k}}
\frac{f(\varepsilon_{+}^{\eta,s_{z}})-f(\varepsilon_{-}^{\eta,s_{z}})}
{(\varepsilon_{+}^{\eta,s_{z}}-\varepsilon_{-}^{\eta,s_{z}})^{2}}
\nonumber \\
&\times&\langle
\Psi^{\eta,s_{z}}_{-}|v_{y}|\Psi^{\eta,s_{z}}_{+}\rangle \langle
\Psi^{\eta,s_{z}}_{+}|v_{x}|\Psi^{\eta,s_{z}}_{-}\rangle,
\label{e4}
\end{eqnarray}
where $A$ is the area of the sample and velocity components can be
obtained from the Hamiltonian and using relation
$v_{k_{i}}=\frac{1}{\hbar}\frac{\partial H}{\partial k_{i}}$.
Furthermore$f(\varepsilon_{\nu}^{\eta,s_{z}})=1/(1+e^{\beta
(\varepsilon_{\nu}^{\eta,s_{z}}-\mu)})$ is Fermi-Dirac
distribution function with $\mu$ being the chemical potential
which at zero temperature is equal to Fermi energy. After
calculating the expectation values of the velocities from Eq.
\ref{e3} and substituting them in Eq. \ref{e4} we get
\begin{eqnarray}
\sigma^{\eta,s_{z}}_{xy}=\eta\frac{e^{2}v_{F}^{2}\hbar}{8\pi}\int
k dk \frac{\Delta_{\eta,s_{z}}}{\chi_{\eta,s_{z}}^{3}}
(f(\varepsilon_{+}^{\eta,s_{z}})-f(\varepsilon_{-}^{\eta,s_{z}})).
\label{e5}
\end{eqnarray}
We restrict our consideration to zero temperature. So we can solve
this equation analytically and obtain a general relation for
$\sigma^{\eta,s_{z}}_{xy}$ for all arbitrary values of exchange
field, $M$, and Fermi energy, $\mu$ which is
\begin{eqnarray}
\sigma^{\eta,s_{z}}_{xy}=-\eta\frac{e^{2}}{4\pi
\hbar}sgn(\Delta_{\eta,s_{z}}), \label{e6}
\end{eqnarray}
when $|\mu+s_{z}M|<|\Delta_{\eta,s_{z}}|$ and
\begin{eqnarray}
\sigma^{\eta,s_{z}}_{xy}=-\eta\frac{e^{2}}{4\pi
\hbar}\frac{\Delta_{\eta,s_{z}}}{|\mu+s_{z}M|}, \label{e7}
\end{eqnarray}
when $|\mu+s_{z}M|>|\Delta_{\eta,s_{z}}|$. These equations are the
main result of this paper. Here sgn(x) is sign function which is 1
for $x>0$, 0 for $x=0$ and -1 for $x<0$. The conditions
$|\mu+s_{z}M|<|\Delta_{\eta,s_{z}}|$ and
$|\mu+s_{z}M|>|\Delta_{\eta,s_{z}}|$ determine boundaries which
separate different phase states. This can be examined by
calculating the spin- and valley-Hall conductivity which are
defined\cite{Vargiamidis1,Kane1,Sinitsyn1,Li1} as
$\sigma^{s}_{xy}=\frac{\hbar}{2e}\sum_{\eta,s_{z}}s_{z}\sigma^{\eta,s_{z}}_{xy}$
and
$\sigma^{v}_{xy}=\frac{1}{2e}\sum_{\eta,s_{z}}\eta\sigma^{\eta,s_{z}}_{xy}$
respectively. This will be explained further in the next section
where we present our results.

\section{Results and discussions}

In this section we present our results. First we examine our
general result in the zero limit of $M$ and $\mu$. When $M=0$ and
$\mu=0$, only $|\mu+s_{z}M|<|\Delta_{\eta,s_{z}}|$ is satisfied,
so we have
$\sigma^{\eta,s_{z}}_{xy}=-\eta\frac{e^{2}}{2h}sgn(\Delta_{\eta,s_{z}})$.
This yields
\begin{eqnarray}
\sigma^{s}_{xy}=-\frac{e}{4\pi}[sgn(\Delta_{+,+})-sgn(\Delta_{+,-})],
\label{e8}
\end{eqnarray}
\begin{eqnarray}
\sigma^{v}_{xy}=-\frac{e}{4\pi \hbar
}[sgn(\Delta_{+,+})+sgn(\Delta_{+,-})], \label{e9}
\end{eqnarray}
for DC spin- and valley-Hall conductivity of a silicene sheet, if
an electric field applied perpendicular to its plane. When
$0\leq|\Delta_{z}|<\Delta_{so}$ these equations yield
$\sigma^{s}_{xy}=-\frac{e}{2\pi}$ and $\sigma^{v}_{xy}=0$,
indicating an intrinsic quantized spin Hall conductivity beside a
zero valley Hall conductivity. This regime, as mentioned in
previous works\cite{Ezawa2,Dyrda1}, corresponds to a topological
insulating (TI) phase characterized by a quantized nonzero
spin-Hall conductivity which arises from the presence of gapless
helical edge mode. When $|\Delta_{z}|$ becomes equal to
$\Delta_{so}$, we have $\sigma^{s}_{xy}=-\frac{e}{4\pi}$ and
$\sigma^{v}_{xy}=-\frac{e}{4\pi \hbar}$. In this regime silicene
is a spin valley polarized metal (SVPM). If the electric filed
increases further such that $\Delta_{so}<|\Delta_{z}|$, the spin
and valley Hall conductivity become $\sigma^{s}_{xy}=0$ and
$\sigma^{v}_{xy}=-\frac{e}{2\pi \hbar}$ corresponding to a
intrinsic quantum valley Hall effect. As mentioned\cite{Ezawa2},
in this regime silicene is a band insulator. These results
indicate an electrically tunable phase transition from a
topological insulator to a spin valley polarized metal and then to
a band insulator, as the electric field increases. It is evident
that our results are in agreement with the similar results
obtained for the DC spin- and valley-Hall conductivity of silicene
in the previous works\cite{Dyrda1,Tahir2,Tabert2}.

In the doped case, when the Fermi level locate inside the gap,
obtained results for $\sigma^{s}_{xy}$ and $\sigma^{v}_{xy}$ are
similar to those of the undoped case. Then, if an external
vertical voltage is applied and increases, a phase transition from
a topological insulator to a metal and then to a band insulator
occurs\cite{Dyrda1}. The boundary conditions which limit different
phases are represented by
$-\Delta_{z}-\Delta_{so}<\mu<\Delta_{z}+\Delta_{so}$ and
$\Delta_{z}-\Delta_{so}<\mu<-\Delta_{z}+\Delta_{so}$ lines for
topological insulating phase and
$-\Delta_{z}+\Delta_{so}<\mu<\Delta_{z}-\Delta_{so}$ or
$\Delta_{z}+\Delta_{so}<\mu<-\Delta_{z}-\Delta_{so}$ lines for
band insulating phase. If the chemical potential increases further
such that the Fermi level crosses the lower conduction (upper
valance) band, the results become
\begin{eqnarray}
\sigma^{s}_{xy}=-\frac{e}{4\pi}[\frac{\Delta_{+,+}}{|\mu|}+1]~~,
~~\sigma^{v}_{xy}=-\frac{e}{4\pi
\hbar}[\frac{\Delta_{+,+}}{|\mu|}-1] \label{e10}
\end{eqnarray}
Notice that in this case, unlike the udoped case, silicene shows a
nonzero DC response for both spin- and valley-Hall conductivity
which are controlled by $\Delta_{so}$ and $\Delta_{z}$. While when
the Fermi level crosses both conduction (valance) bands, DC spin-
and valley Hall conductivity become
\begin{eqnarray}
\sigma^{s}_{xy}=-\frac{e}{4\pi}\frac{\Delta_{so}}{|\mu|}~~,
~~\sigma^{v}_{xy}=\frac{e}{4\pi \hbar}\frac{\Delta_{z}}{|\mu|}.
\label{e12}
\end{eqnarray}
On can see that in this case the DC spin-Hall conductivity is an
intrinsic property and is only controlled by $\Delta_{so}$,
whereas the valley-Hall conductivity is arising from and tuned by
the applied vertical voltage. Moreover, the direction of the
valley Hall conductivity changes by inverting the direction of the
applied electric filed while the direction of the spin Hall
conductivity remains unchanged. These results for the doped case
are in agreement with the results reported in the previous
works\cite{Dyrda1,Tahir2,Tabert2}. All these results have been
summarized in Figs.\ref{f2} and \ref{f3} where we have shown the
spin and valley Hall conductivity of a electron/hole doped
silicene as a function of the vertical electric filed and the
Fermi energy.

It is evident that in these cases, due to the symmetry of the band
structure with respect to interchanging the valley and spin index,
there is no spin or valley polarization.

Furthermore, the DC spin- and valley-Hall conductivity of a
undoped ferromagnetic silicene can be obtained by making use of
Eqs. \ref{e6} and \ref{e7}. When the exchange field is less than
the minimum gap, at low external vertical voltage, silicene shows
a quantum spin Hall effect with an intrinsic quantized Hall
conductivity, $\sigma^{s}_{xy}=-\frac{e}{2\pi}$. By increasing the
external vertical voltage silicene becomes metal with nonzero
charge- spin- and valley-Hall conductivity. Then at high external
vertical voltages a phase transition to a conventional band
insulator (characterized by zero charge and spin-Hall
conductivities and a nonzero quantized valley-Hall conductivity,
$\sigma^{v}_{xy}=-\frac{e}{2\pi \hbar}$) occurs. The necessary
conditions to realize a quantum spin Hall effect are represented
by $-\Delta_{z}-\Delta_{so}<\mu<\Delta_{z}+\Delta_{so}$ and
$\Delta_{z}-\Delta_{so}<\mu<-\Delta_{z}+\Delta_{so}$ lines. While
the similar conditions for a quantum valley Hall effect are
$-\Delta_{z}+\Delta_{so}<M<\Delta_{z}-\Delta_{so}$ or
$\Delta_{z}+\Delta_{so}<M<-\Delta_{z}-\Delta_{so}$. Furthermore,
when the Fermi level crosses the lower conduction (upper valance)
band, our results for the DC spin and valley Hall conductivity
become
\begin{eqnarray}
\sigma^{s}_{xy}=-\frac{e}{4\pi}[\frac{\Delta_{+,+}}{|M|}+1] ~~ ,
~~ \sigma^{v}_{xy}=-\frac{e}{4\pi
\hbar}[\frac{\Delta_{+,+}}{|M|}-1] \label{e10}
\end{eqnarray}
Then if the exchange filed increases further such that the Fermi
level crosses both conduction (valance) bands, DC spin- and valley
Hall conductivity become
\begin{eqnarray}
\sigma^{s}_{xy}=-\frac{e}{4\pi}\frac{\Delta_{so}}{|M|}
~~,~~\sigma^{v}_{xy}=\frac{e}{4\pi \hbar}\frac{\Delta_{z}}{|M|}.
\label{e12}
\end{eqnarray}
Note, if the Fermi level crosses both conduction (valance) bands,
the DC spin and valley Hall conductivity in a ferromagnetic
silicene are intrinsic properties which are controlled only by
$\Delta_{so}$ and $\Delta_{z}$ respectively. Moreover, one can see
that in a undoped ferromagnetic silicene similar to the previous
case, due to the symmetry of the band structure with respect to
interchanging the valley and spin index, any spin/valley polarized
transport can not be attained. To achieve a valley or spin
polarized transport, one can populate spin or valley states
differentially in one valley or spin state. This can be attained,
as explained in the following, in a doped ferromagnetic silicene.

Our results for the spin- and valley Hall conductivity in a doped
ferromagnetic silicene have been summarized in Fig.\ref{f4} and
Fig. \ref{f5}. One can see that the region, in which a quantized
spin or valley Hall effect occur, becomes limited when the induced
exchange field increases. The necessary conditions to realize a
quantum spin Hall effect are represented by
$-\Delta_{z}-\Delta_{so}<\mu<\Delta_{z}+\Delta_{so}$ and
$\Delta_{z}-\Delta_{so}<\mu<-\Delta_{z}+\Delta_{so}$ lines. While
the similar conditions for a quantum valley Hall effect are
$-\Delta_{z}+\Delta_{so}<M<\Delta_{z}-\Delta_{so}$ or
$\Delta_{z}+\Delta_{so}<M<-\Delta_{z}-\Delta_{so}$.

Moreover, as mentioned above, a partial or fully spin/valley
polarized transport can be captured in a doped ferromagnetic
silicene. This can be seen in Fig.\ref{f6} and Fig. \ref{f7}. In
Fig.\ref{f6} we have shown our results for
$\sigma^{\uparrow}_{xy}=\sigma^{K\uparrow}_{xy}
+\sigma^{K^{'}\uparrow}_{xy}$ and
$\sigma^{\downarrow}_{xy}=\sigma^{K\downarrow}_{xy}
+\sigma^{K^{'}\downarrow}_{xy}$. The regions, in which
$\sigma^{\uparrow}_{xy}(\frac{e}{4\pi})$ or
$\sigma^{\downarrow}_{xy}(\frac{e}{4\pi})$ is zero, determine the
necessary conditions to realize a fully spin polarized transport.
These regions have been shown in Fig. \ref{f8}. Furthermore, in
Fig. \ref{f7} the plots of
$\sigma^{K}_{xy}=\sigma^{K\uparrow}_{xy}
+\sigma^{K\downarrow}_{xy}$ and
$\sigma^{K^{'}}_{xy}=\sigma^{K^{'}\uparrow}_{xy}
+\sigma^{K^{'}\downarrow}_{xy}$, as a function of the vertical
electric filed and the chemical potential, have been shown. The
regions, in which a fully valley polarized transport can be
detected, have been shown in Fig. \ref{f8}.

\section{Summary and conclusions}
\label{sec:03}

In summary we studied the intrinsic DC valley and spin Hall
conductivity in a ferromagnetic silicene, exploring a fully spin
or valley polarized transport. First we calculated its eigenvalues
and eigenfunctions. Then, by making use of the Kubo formula, we
derived a general relation for the spin and valley Hall
conductivity of the ferromagnetic silicene in the presence of
finite doping and an electric filed applied perpendicular to its
plane. We examined our result by reproducing the results of the
previous works in the zero limit of the exchange field. Moreover,
we calculated the DC spin and valley conductivity of a doped
ferromagnetic silicene. Finally we used our general result to
determine the necessary conditions for realizing a fully spin or
valley polarized transport.

%

%
%
\newpage
\begin{figure}
\begin{center}
\includegraphics[width=15cm,angle=0]{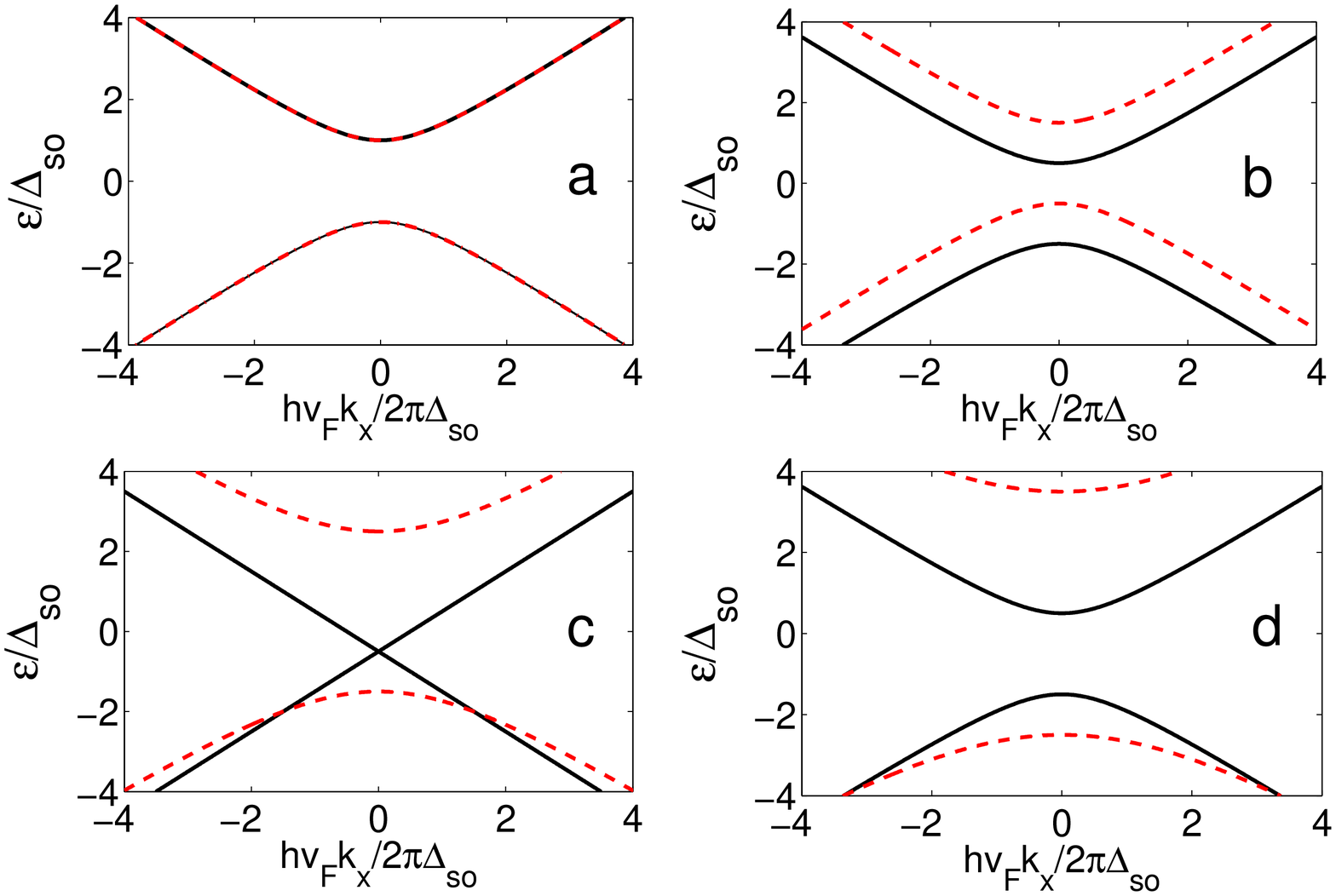}
\caption{The energy spectrum of (a) silicene, and ferromagnetic
silicene with $M=\Delta_{so}/2$ when (b) $\Delta_{z}=0$, (c)
 $\Delta_{z}=\Delta_{so}$ and (d) $\Delta_{z}=2\Delta_{so}$.}\label{f1}
\end{center}
\end{figure}
\begin{figure}
\begin{center}
\includegraphics[width=15cm,angle=0]{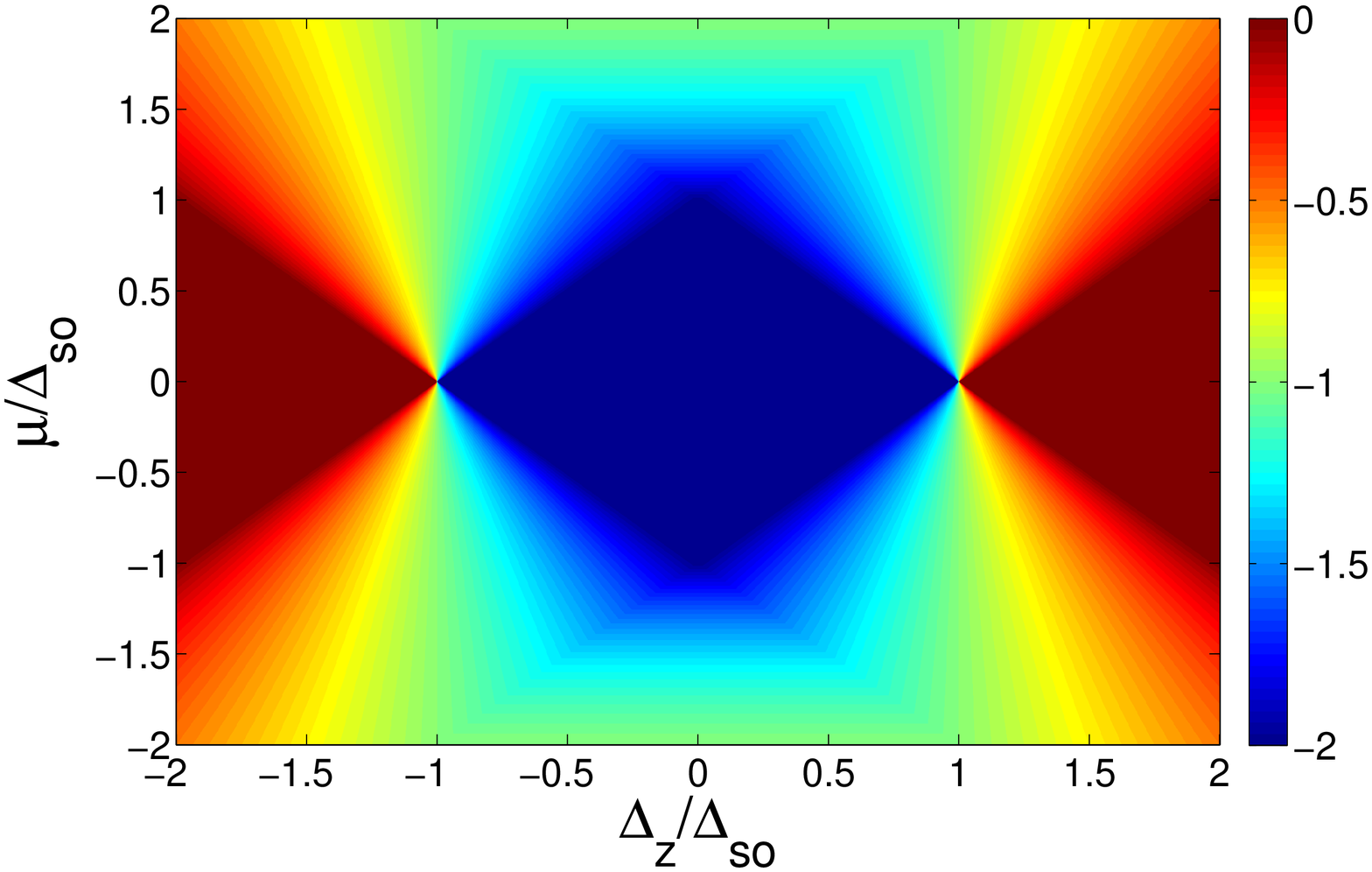}
\caption{The transverse spin Hall conductivity,
$\sigma^{s}_{xy}(\frac{e}{4\pi})$, in a doped silicene as a
function of the vertical electric filed and the chemical
potential.}\label{f2}
\end{center}
\end{figure}
\begin{figure}
\begin{center}
\includegraphics[width=15cm,angle=0]{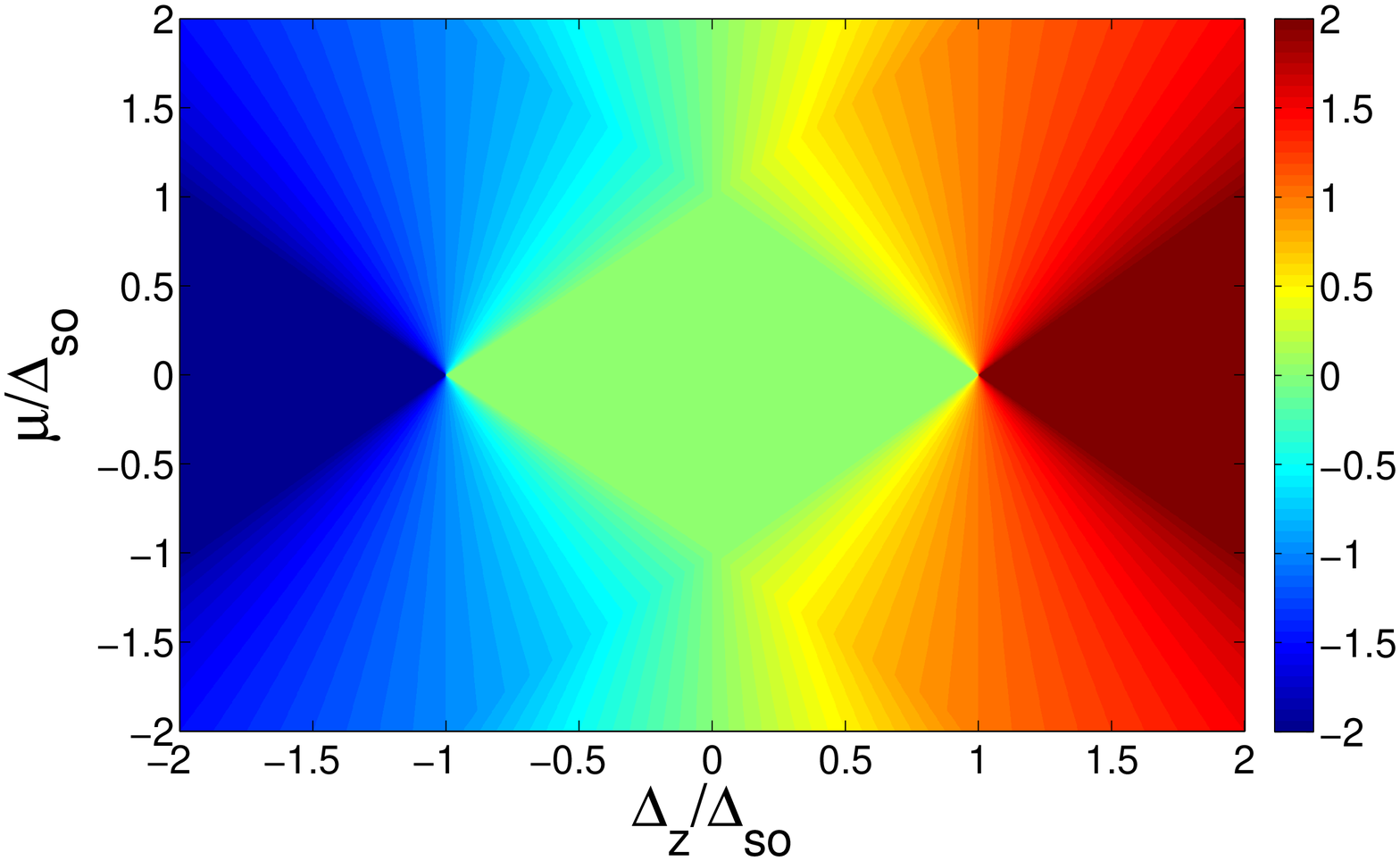}
\caption{The transverse valley Hall conductivity,
$\sigma^{v}_{xy}(\frac{e}{4\pi\hbar})$, in a doped silicene as a
function of the vertical electric filed and the chemical
potential.}\label{f3}
\end{center}
\end{figure}
\begin{figure}
\begin{center}
\includegraphics[width=15cm,angle=0]{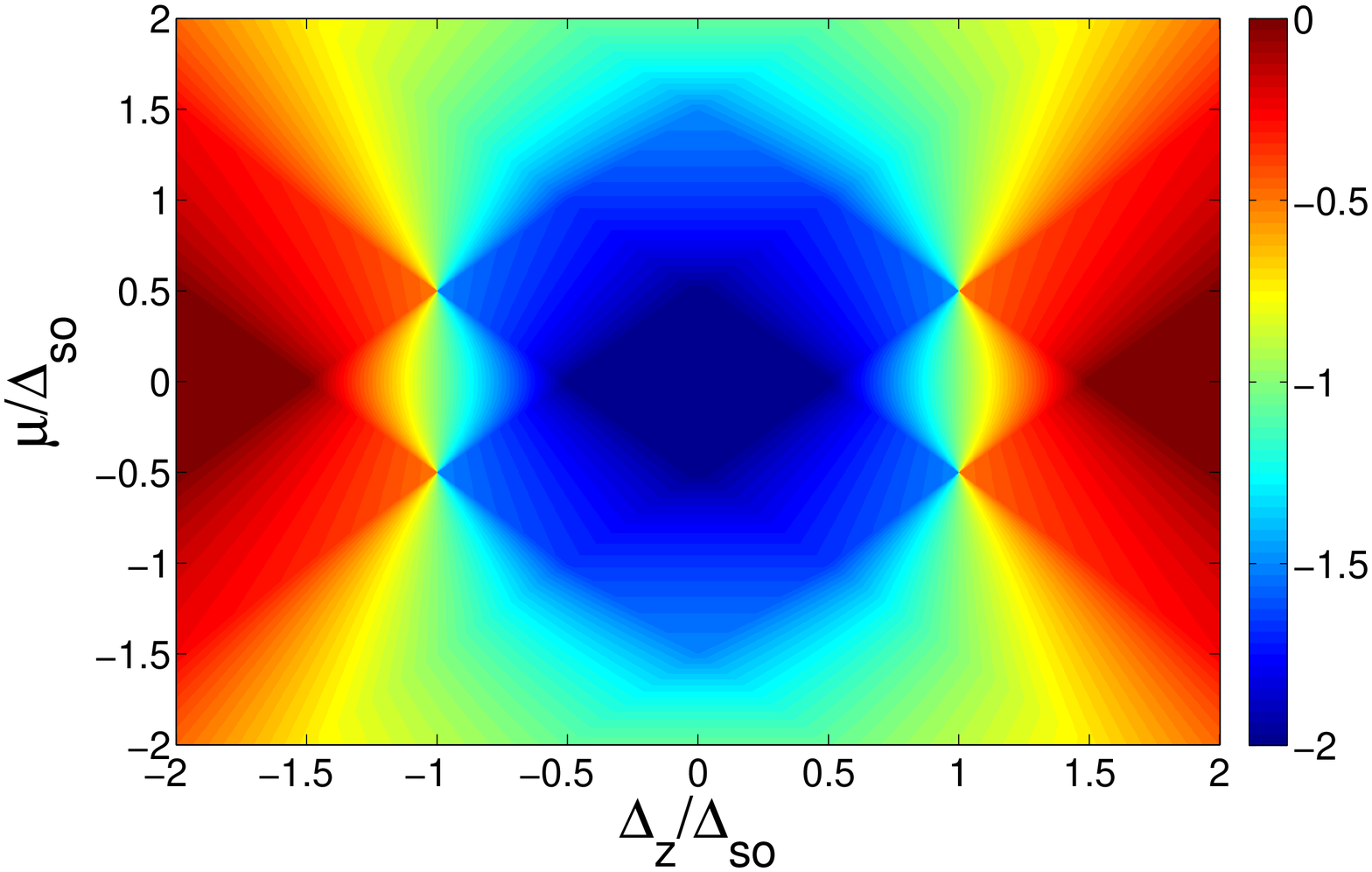}
\caption{The transverse spin Hall conductivity,
$\sigma^{s}_{xy}(\frac{e}{4\pi})$, in a doped ferromagnetic
silicene with $M=\Delta_{so}/2$, as a function of the vertical
electric filed and the chemical potential.}\label{f4}
\end{center}
\end{figure}
\begin{figure}
\begin{center}
\includegraphics[width=15cm,angle=0]{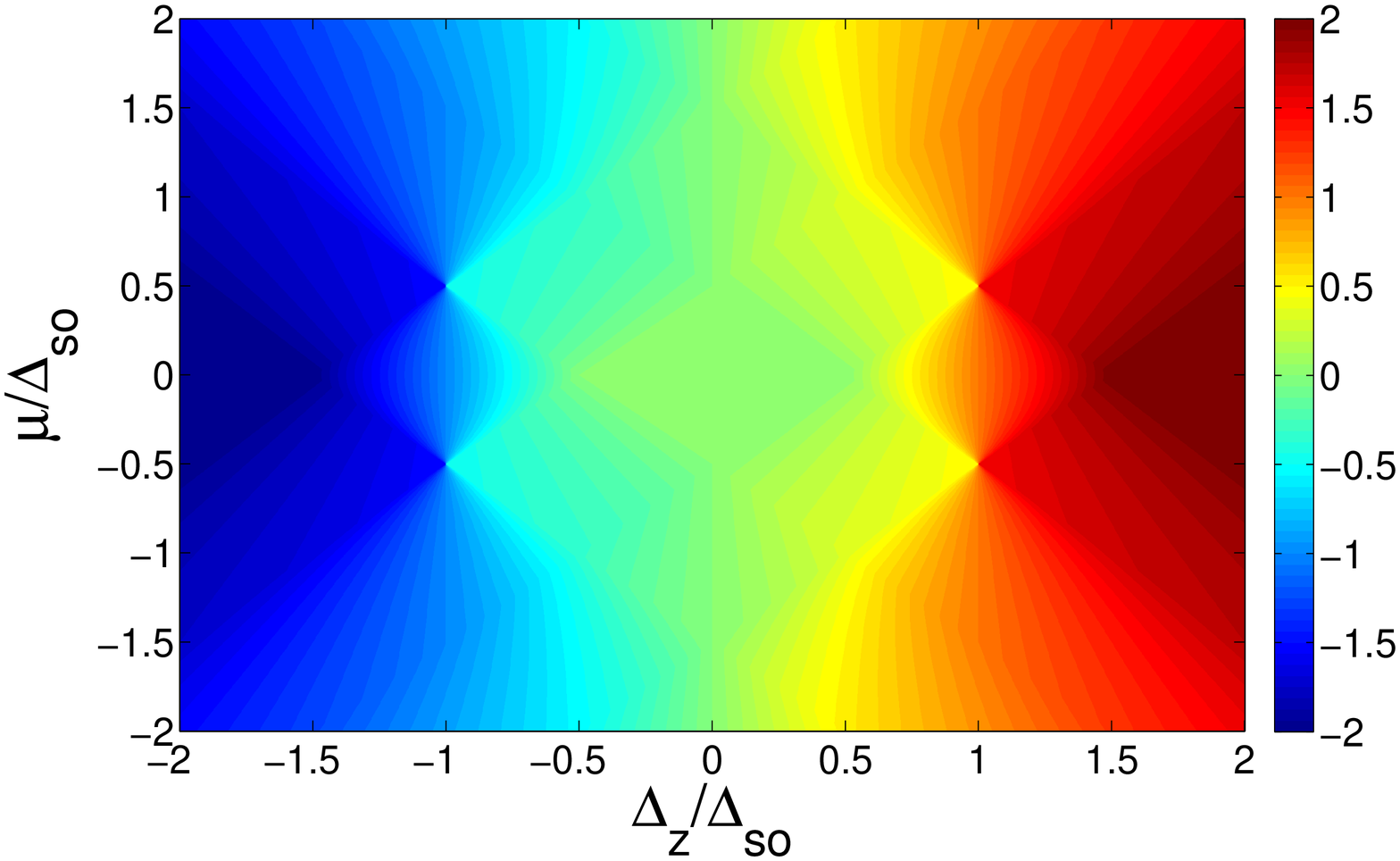}
\caption{The transverse valley Hall conductivity,
$\sigma^{v}_{xy}(\frac{e}{4\pi\hbar})$, in a doped ferromagnetic
silicene with $M=\Delta_{so}/2$, as a function of the vertical
electric filed and the chemical potential.}\label{f5}
\end{center}
\end{figure}
\begin{figure}
\begin{center}
\includegraphics[width=15cm,height=15cm,angle=0]{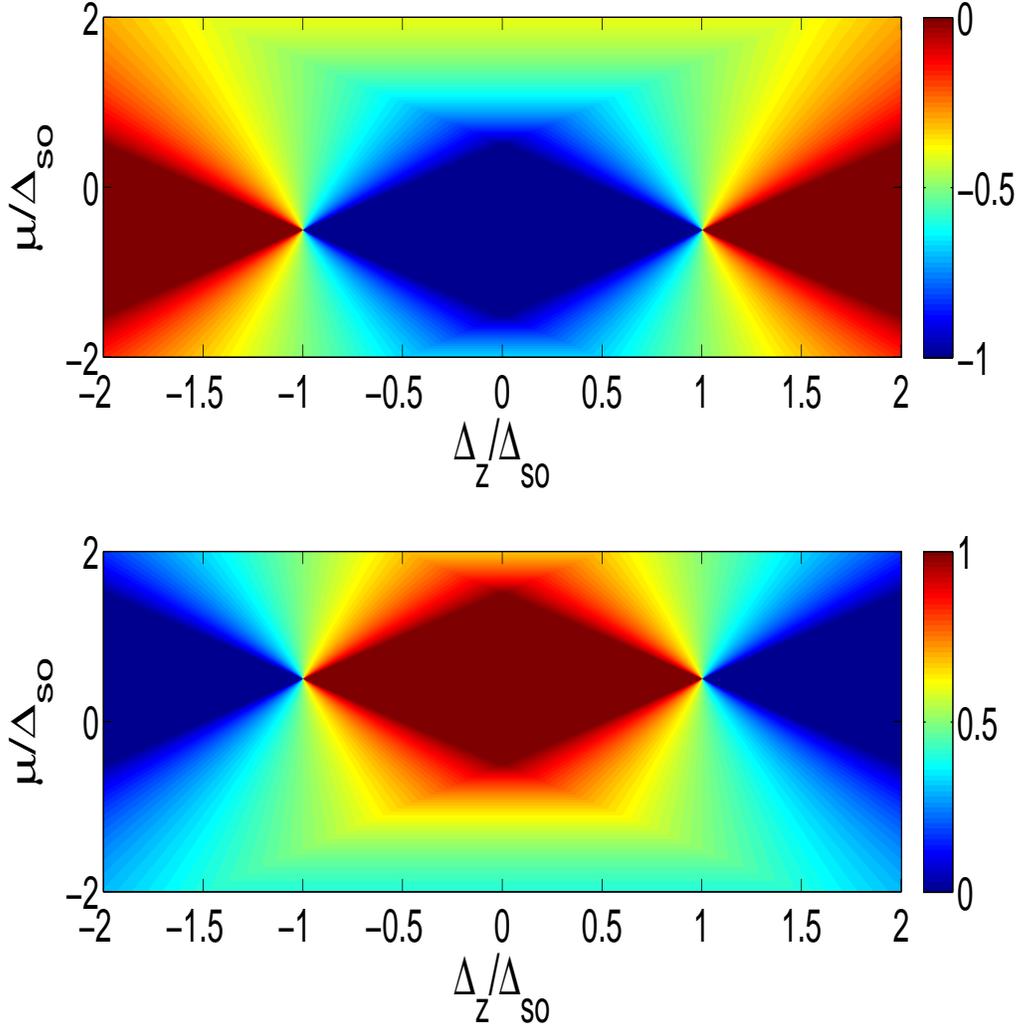}
\caption{Plots of $\sigma^{\uparrow}_{xy}(\frac{e}{4\pi})$ (top)
and $\sigma^{\downarrow}_{xy}(\frac{e}{4\pi})$ (bottom), in a
doped ferromagnetic silicene with $M=\Delta_{so}/2$, as a function
of the vertical electric filed and the chemical potential. The
regions in which $\sigma^{\uparrow}_{xy}(\frac{e}{4\pi})$ or
$\sigma^{\downarrow}_{xy}(\frac{e}{4\pi})$ is zero determine the
necessary conditions to realize a fully spin polarized
transport.}\label{f6}
\end{center}
\end{figure}
\begin{figure}
\begin{center}
\includegraphics[width=15cm,height=15cm,angle=0]{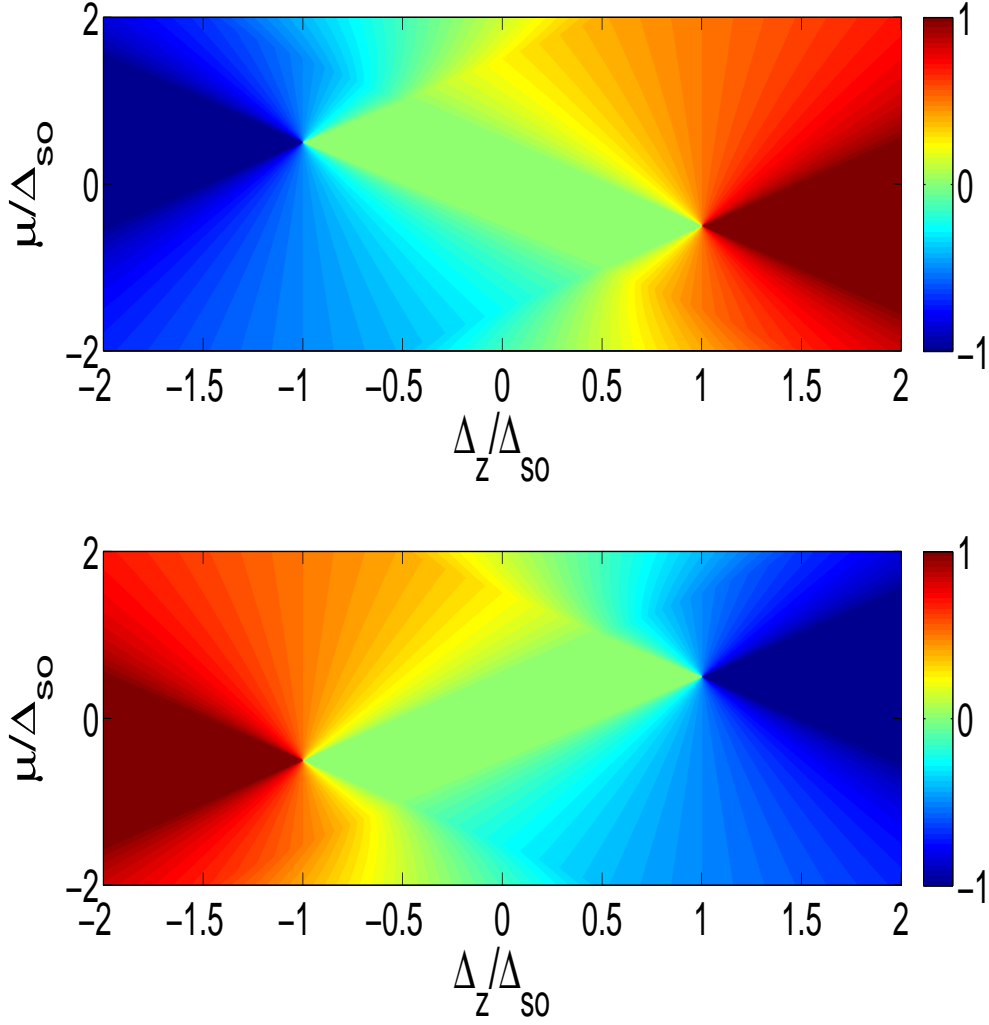}
\caption{Plots of $\sigma^{K}_{xy}(\frac{e}{4\pi\hbar})$ (top) and
$\sigma^{K^{'}}_{xy}(\frac{e}{4\pi\hbar})$ (bottom), in a doped
ferromagnetic silicene with $M=\Delta_{so}/2$, as a function of
the vertical electric filed and the chemical potential. The
regions in which $\sigma^{K}_{xy}(\frac{e}{4\pi\hbar})$ or
$\sigma^{K^{'}}_{xy}(\frac{e}{4\pi\hbar})$ is zero determine the
necessary conditions to realize a fully valley polarized
transport.}\label{f7}
\end{center}
\end{figure}
\begin{figure}
\begin{center}
\includegraphics[width=15cm,angle=0]{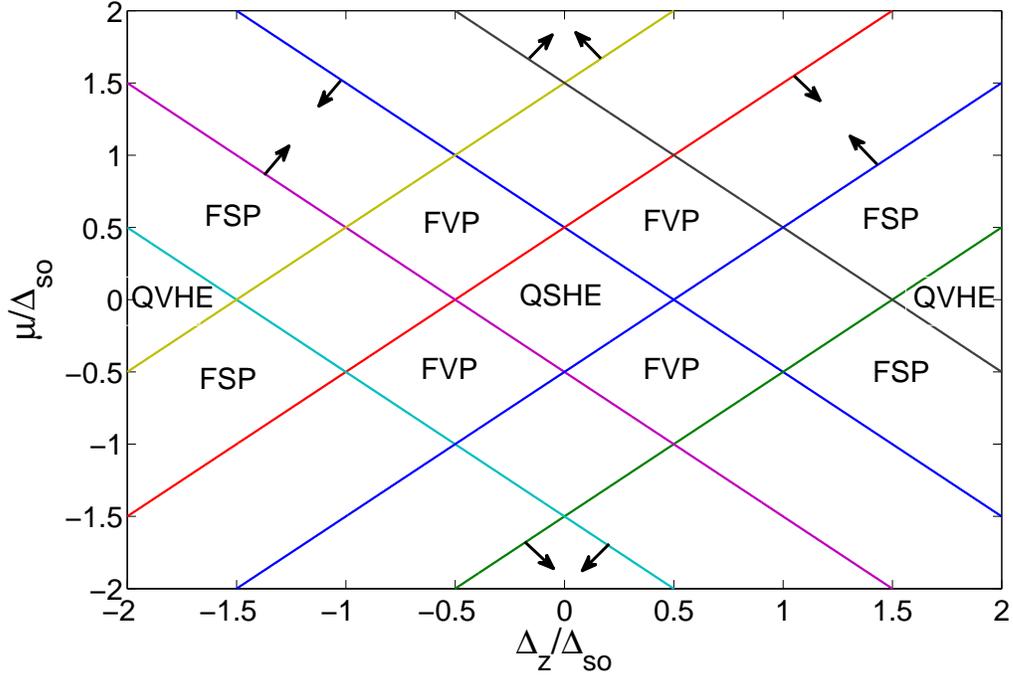}
\caption{Regions which determines the necessary conditions to
realize a fully spin polarized (FSP) or a fully valley polarized
(FVP) transport and also quantum spin Hall effect (QSHE) and
quantum valley Hall effect (QVHE). Arrows show how the regions
increase or decrease by increasing the exchange field. }\label{f8}
\end{center}
\end{figure}
\end{document}